\documentclass{article}

\usepackage{arxiv}
\usepackage{graphicx}
\usepackage{subcaption}

\usepackage[utf8]{inputenc} 
\usepackage[T1]{fontenc}    
\usepackage{hyperref}       
\usepackage{url}            
\usepackage{booktabs}       
\usepackage{amsfonts}       
\usepackage{nicefrac}       
\usepackage{microtype}      
\usepackage{lipsum}
\usepackage{graphicx}
\graphicspath{ {./images/} }

\title{Towards the Understanding of Receptivity and Affect in EMAs using Physiological based Machine Learning Method: Analysis of Receptivity and Affect}

\author{
 Zachary D King \\
  Department of Electrical \\and Computer Engineering\\
  Rice University\\
  Houston, TX, 77005 \\
   \\
   \And
Han Yu \\
  Department of Electrical \\and Computer Engineering\\
  Rice University\\
  Houston, TX, 77005 \\
   \\
  \And
 Thomas Vaessen \\
  Center For Contextual Psychiatry\\
  KU Leuven\\
  Leuven, Belgium \\
  \\
  \And
 Iniz Myin-Germeys \\
  Center For Contextual Psychiatry\\
  KU Leuven\\
  Leuven, Belgium \\
  \\
  \And
 Akane Sano \\
  Department of Electrical \\and Computer Engineering\\
  Rice University\\
  Houston, TX, 77005 \\
}

\begin{document}
\maketitle
\begin{abstract}
\textbf{Background:} As mobile health (mHealth) studies become increasingly productive due to the advancements in wearable and mobile sensor technology, our ability to monitor and model human behavior will be constrained by participant receptivity. Many health constructs are dependent on subjective responses, and without such responses, researchers are left with little to no ground truth to accompany our ever-growing biobehavioral data. This issue can significantly impact the quality of a study, particularly for populations known to exhibit lower compliance rates. To address this challenge, researchers have proposed innovative approaches using machine learning and sensor data to modify the timing and delivery of surveys. However, an overarching concern is the potential introduction of biases or unintended influences on participants' responses when implementing new survey delivery methods.

\textbf{Objective:} We examine the factors that affect participants’ receptivity to ecological momentary assessments (EMA) in a 10-day wearable and EMA-based emotional state sensing mHealth study. We study the physiological relationships indicative of receptivity and affect while also analyzing the interaction between the two constructs. We demonstrate the potential impact of an ML-based EMA delivery system (using receptivity as the predictor variable) on the participants' reported emotional state. 

\textbf{Methods:} We collected data from 45 healthy participants wearing two devices measuring electrodermal activity, acceleration, electrocardiography, and skin temperature while answering 10 EMAs daily containing questions about perceived mood. Due to the nature of our constructs, we can only obtain ground truth measures for both affect and receptivity during responses. Therefore, we utilized unsupervised and supervised machine learning methods to infer affect when a participant did not respond. Our unsupervised method used k-means clustering to find the relationship between physiology and receptivity and then inferred the emotional state during non-responses. For the supervised learning method, we primarily used Random Forest (RF) and Neural Networks (NN) to predict affect of unlabeled data points as well as receptivity. 

\textbf{Results:} Our findings showed using a receptivity model to trigger EMAs decreased the reported negative affect by more than 3 points or 0.29 standard deviations in our self-reported affect measure scored between 13 and 91. The findings also showed a bimodal distribution of our predicted affect during non-responses. This would indicate that this system would initiate EMAs more commonly during states of higher positive emotions.

\textbf{Conclusions:} Our results showed a clear relationship between affect and receptivity. This relationship can affect the efficacy of a mHealth study, particularly those studies that employ a machine learning algorithm to trigger EMAs. Therefore, we propose a smart trigger that promotes EMA receptivity without influencing affect during sampled time points as future work.

\end{abstract}


\section{Introduction}

\subsection{User Engagement in Mobile Health Systems}

Mobile health (mHealth) technologies continue to grow within the healthcare sector and are imperative in the precision medicine initiative. mHealth can provide beneficial interactions between healthcare providers and patients outside clinical settings. An engaged and responsive user base in any mHealth system is vital in maximizing the knowledge that researchers and providers acquire. Mental health research mainly depends on active users because investigators rely on participant survey responses to establish ground truth. Researchers can only adequately interpret the relationships between physiology and psychological state with a population compliant with sensors and surveys. Evaluating a health construct is only possible with highly receptive participants in these mHealth studies. 

Here, we discuss two forms of interaction between participants and mHealth systems: Ecological Momentary Assessments (EMA) and Just-in-Time Interventions (JITI). EMAs gather in-situ data from users in real time. EMAs are commonly used in mHealth studies as they allow researchers to prompt subjects regularly throughout the day \cite{stone1994ecological}. In the case of mHealth studies focusing on psychological states, EMAs enable users to report their momentary symptoms or context in a natural environment, often using smartphones due to their accessibility. JITI is a method that allows investigators to send interventions as needed. The adaptive version of the JITI (JITAI) uses incoming information (physiological, contextual, or psychological markers) as context to determine when an intervention is needed \cite{morris2005embedded}. Researchers have been working to enhance the efficiency of these interactions. As we mentioned previously, this effort is crucial because ineffective interactions in a mHealth study can have significant effects on outcomes. Failing to collect EMA responses may impede researchers' ability to identify real-world measures of health behaviors, and without participants receiving or engaging in JITIs, researchers may find it challenging to measure the efficacy of the intervention.


\subsection{Improving EMA Receptivity }

To enhance compliance with EMAs and JITIs, it is imperative to gain a comprehensive understanding of the factors that influence participant adherence. Ho et al. \cite{ho2005using} described 11 factors that influence a person's interuptability (willingness to follow through if notified or interrupted). These factors encompass contextual aspects, such as social engagement, ongoing activities, future schedule, and emotional state, as well as message-related attributes, including frequency, complexity, modality, and utility. 

Currently, many researchers have reduced interuptability by altering message-related attributes, often involving strategies like reducing the complexity or frequency of an EMA or increasing the incentives for a response \cite{intille2016muema, king2019micro}. Reducing the size of the instrument relieves some of the burden associated with answering an EMA \cite{ponnada2017microinteraction}. This is done by excluding redundant questions or choosing a less complex instrument. The Perceived Stress Scale (PSS) \cite{cohen1988perceived} was initially a 14-item question set. Still, after some statistical analysis, researchers found that a 10-item instrument was sufficient for measuring stress. Another factor affecting receptivity is the frequency at which users are sampled. In two separate reviews, researchers demonstrated conflicting findings on the effects of frequency on EMA compliance \cite{wen2017compliance, jones2019compliance}. These conflicting results can be attributed to the varying populations the authors focused on and the many other factors that play a role in EMA compliance. The third method for improving receptivity rates is increasing the incentives based on EMA compliance. However, this method can be costly and seen as exploitative, especially when dealing with vulnerable populations.


An emerging method for improving receptivity rates is the use of machine learning. This can be achieved by utilizing wearable data to predict the likelihood of a response, which can help deliver EMAs that mitigate interuptability.  Mishra et al. used machine learning models built from previously collected data to improve the receptivity of a JITAI by contacting users at points where they are more likely to be receptive \cite{mishra2021detecting}. The study showed a difference of over 38\% in receptivity rates between an ML-based static model (using data collected previously) and a control model (using a set schedule) to distribute EMAs. Mishra et al. built a model for predicting the optimal time to send an EMA \cite{mishra2017investigating}. Their results demonstrated that a model built from contextual cues like activity, audio, conversation, and location could significantly outperform a baseline model (prediction based on the proportion of responded EMAs). Researchers have also shown that contextual cues, including location \cite{morrison2017effect, pielot2017beyond}, personality traits \cite{kunzler2019exploring, mehrotra2016my}, physical activity \cite{kunzler2019exploring, mehrotra2015designing}, and time of day \cite{bidargaddi2018prompt}, have influenced a participant's willingness to respond to regular surveys. Together, these methods can predict and respond to the unobserved contextual aspects of an interruption, thus offering a more holistic approach to addressing participant engagement. However, a system that reacts to these contextual aspects may have unintentional effects on the response of the user. For instance, emotional state is an underlying factor that affects receptivity. A model designed to initiate EMAs when a participant is most likely to respond favors prompting users when experiencing positive emotions. As a result, this approach could influence the reported emotional state during each prompt, potentially making it challenging to collect subjective responses during negative emotions. Understanding the influence that machine learning-based EMA triggers exert on these underlying receptivity factors allows us to incorporate additional variables into an algorithm. Integrating predicted affect in the decision-making of an ML-based EMA trigger will ensure that participants receive prompts across a broad spectrum of emotions. 




\subsection{Relationship between Affect and EMA Receptivity}

Clark et al. \cite{clark1988mood} described how positive and negative affect can influence participation in activities of daily living (ADL). Their results show differences in expected mean across many social activities, with reported positive affect having more significance in differentiating the two groups. Similarly, research has also demonstrated a negative relationship between students' emotional state and academic achievement \cite{duchesne2008trajectories, valiente2012linking}. While none of these studies demonstrate the relationship between affect and EMA receptivity during mHealth studies, they all demonstrate the effect of emotional state on a participant's general ability to engage in normal ADL. 

Several authors have examined the effect of emotional state on EMA adherence by using the preceding response as a gauge of affect during instances of non-response. Murray et al. conducted a study (N = 261) demonstrating that negative affect and stress reduce the chance of a response during the next prompt \cite{murray2023prompt}. Other researchers have expanded on this by examining various contextual cues within an EMA that precede instances of non-response. The authors found that variables like medication use, activity, battery life, and being away from home negatively impacted the compliance of the following EMA \cite{ponnada2022contextual, rintala2020momentary}. This work contributes to understanding how affect can influence participants' response behavior but falls short in providing real-time explanations for the absence of responses. 
Alternatively, real-time explanations for receptivity can be derived through passive sensing and machine learning. Leveraging these explanations allows for delivering EMAs at moments of heightened receptivity, guided by current contextual and physiological factors.  


\subsection{Objectives and Hypothesis}

In this study, we aim to analyze the relationship between participant EMA receptivity and affect in a 10-day wearable and EMA-based affect sensing study (N = 45). We hypothesize a relationship exists between EMA receptivity and affect in mental health-related mHealth studies. We can establish the relationship between emotions when participants respond. However, to investigate this connection during non-responses, we need to infer affect when a participant fails to provide a response. Therefore, we implement machine learning models for identifying receptive time points and predicting emotional states. This allows us to determine if there is a statistically significant difference in emotions between responses and non-responses. If this relationship exists, and the likelihood of a response is dependent on emotional state, it would bias the outcome of a machine learning-based EMA delivery mechanism. 

\section{Methods}
\subsection{Ethics Approval}

Ethical approval was granted by the Sociaal-Maatschappelijke Etische Commissie (SMEC) of KU Leuven (G-2018 09 1339) \cite{de2023daily}.

\subsection{Data Collection}
The study included 45 healthy adult participants in Leuven, Belgium. The average age of the subjects was 24.5 years and ranged between 19 and 35 years. Thirty-eight (85\%) of the subjects were female. Participants were recruited via flyers distributed to areas around Leuven. 

\begin{figure}
\begin{center}
\includegraphics[width=0.5\textwidth]{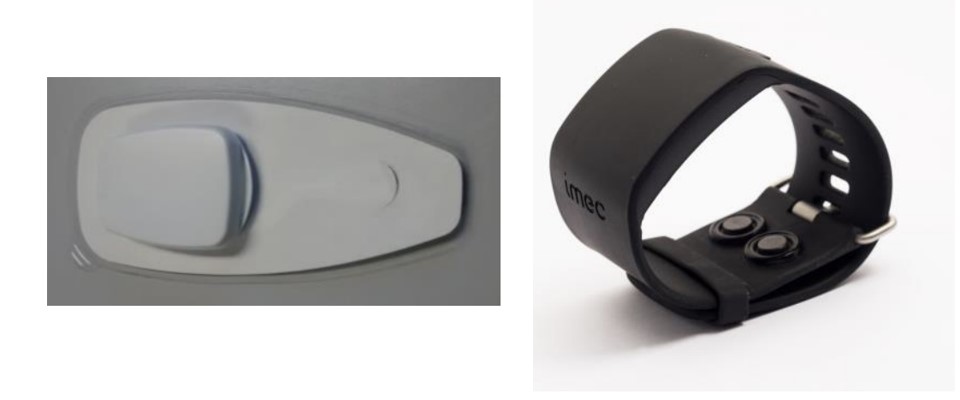}
\end{center}
\caption{Left: chest patch for gathering electrocardiogram (ECG) and acceleration (ACC). Right: wristband for gathering electrodermal
activity (EDA), skin temperature, and ACC.}
\label{fig:sensor}
\end{figure}

The study lasted for 10 days. The participants wore a sensor suite shown in Figure ~\ref{fig:sensor}, including a chest patch with two electrodes for gathering Electrocardiography (ECG) at 256 Hz. and a wristband for electrodermal activity (EDA) at 256 Hz, skin temperature (ST) at 1 Hz. and acceleration (ACC) at 32 Hz. Participants were allowed to remove the device while they slept and were asked to remove the devices while they bathed or participated in rigorous activity. The sensors had battery life that surpassed the duration of the study, and data was recorded on the device on an SD card.

 Participants were given a research phone, and 10 EMAs were sent to the participants daily at random time points between 15 and 90 minutes apart. EMAs were initiated via text message, and the subject had a specific amount of time to respond to the survey attached to the text message before it closed. The EMAs contained a question set to assess mood \cite{myin2001emotional} in three languages: English, Belgian, and French. In total, there were 13 questions, including 9 negative (worried, stressed, anxious, annoyed, down, restless, tense, under pressure, and ashamed) and 4 positive (relaxed, cheerful, confident, and in-control) affect-related questions. The questions were prefaced with the phrase, 'At the moment, I feel...' followed by a rating scale for each emotion, ranging from 1 (not at all) to 7 (very much). Participants were given 50 cents for each EMA they responded to.

\subsection{EMA Analysis}

Scoring our EMA question set was done by adding the numerical interpretation of the nine negative responses to the inverse (1 is 7 and 7 is 1) of the positive questions. The range of possible scores is between 13 and 91, with higher scores relating to more negative emotions. Due to the low variance in reported positive and negative affect, we utilized a composite score of both positive and negative affect. 

We also analyzed the participants’ response time (time between the notification and onset of EMA) and response rate to EMA. We then investigated the potential for loss of engagement over time, which may lead to reduced participant receptivity. Lack of engagement may impede our capacity to discern the underlying causes of non-responsiveness, particularly when assessing the relationship between affect and receptivity. 

\subsection{EMA Receptivity and Affect Detection Models}

\begin{figure}
 \centering
\includegraphics[width=0.85\textwidth]{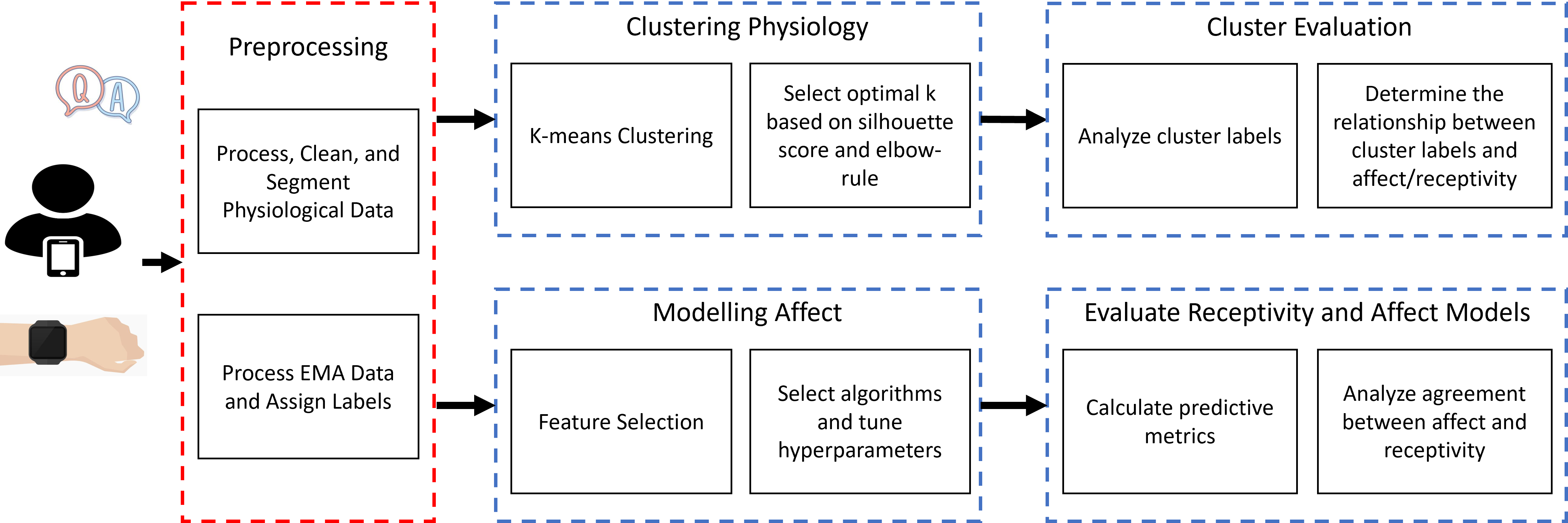}
\caption{Methodology used from the raw signals to our evaluation of the relationship between Affect and Receptivity}
\label{fig:4}
\end{figure}

In the following sections, we discuss the sequential methodology, which encompasses the collection of raw signal data, the subsequent data processing and feature extraction, and the design of machine learning models for inferring two constructs, receptivity and affect. This framework is shown in Figure ~\ref{fig:4}. 

We began by processing our four sets of time series data, ST, ECG, EDA, and ACC. Once we have processed the data, we segmented it and attached labels to each segment based on conditions explained in a later section. Next, we built and tested multiple machine-learning algorithms to infer EMA receptivity and affect, and verified the results using several statistical techniques. 

\subsubsection{Preprocessing}

\subsubsubsection{Time Series Processing}

We began by extracting all the data from the four time series data sets. See features computed for the four sets of the data in Table ~\ref{tab:features}. We used the interquartile range (IQR = Q3 - Q1) to process ST to remove outliers. We used biosppy \cite{biosppy} for ECG to process the data and extract R peaks. Biosppy uses a Bandpass filter with frequencies at 3 Hz and 45 Hz, a sampling rate of 256, and the Hamilton segmentation algorithm to extract R peaks. We then validated the R peaks using an algorithm from Hovesepian et al. \cite{hovsepian2015cstress}; this algorithm uses the criterion beat difference (CBD) based on the Maximum Expected Difference (MED) for a beat and the Minimal Artifact Difference (MAD). We then used HRVanalysis to extract heart rate and heart rate variability (HRV) features such as NN20 and RMSSD \cite{pichot2016hrvanalysis}. We also obtained some frequency and geometric-based features. For EDA, we used the method proposed by \cite{taylor2015automatic} to process and extract statistical and wavelet features. Finally, for ACC, we smoothed the signal by using a 4th order 10Hz low pass Butterworth filter and obtaining an average, then we used a package from \cite{sensormotion} to extract step features. The features we extracted and information on how those features are calculated can be seen in Table ~\ref{tab:features}

\begin{table}
\centering  
\resizebox{0.95\linewidth}{!}{\begin{tabular}{p{0.161\linewidth} | p{0.3\linewidth} | p{0.5\linewidth} | p{0.1\linewidth}}
\hline
\textbf{Signal} &  \textbf{Features} & \textbf{Description} & \textbf{Prior Work}\\
\hline
\textbf{Skin} \textbf{Temperature} \textbf{(ST)} & mean, median, mode, minimum, range, root mean square, zero cross, Kurtosis, skew, IQR 25\textsuperscript{th} and 75\textsuperscript{th} Percentile &Zero-cross here is based on the number of times ST crosses over the mean ST. Kurtosis measures the extremity of the data in the segment and skew is the measure of asymmetry.  & \cite{huynh2021stressnas}\\
\hline
\textbf{Electrocardiogram} \textbf{(ECG)} & mean, median, mode, minimum, range, root mean square, zero cross Kurtosis, skew, IQR 25\textsuperscript{th} and 75\textsuperscript{th} Percentile RMSSD, CVSD, CVNNI SDNN, NNI50, NNI20, PNNI50, PNNI50, low frequency (lf), very low frequency (vlf), high frequency (hf), high/low frequency ratio (hf/lf) &  NN (N-N or R-R interval) indicates the time between heartbeats. NNI20/50 refers to the number of successive intervals that differ by more than 20 or 50 ms. P indicates the proportion of NNI20/50 in the segment. RMSSD is the root mean square of successive differences between heartbeats. CVNNI and CVSD are the coefficients of variation (sdnn/mean) and (rmssd/mean), respectively. Our frequency domain features are based on how much of the signal lies between 0.003 to 0.04 Hz (vlf),  0.04 to 0.15 Hz (lf), 0.15 to 0.40 Hz (hf) & \cite{ha2021wistress},  \cite{he2017emotion}, \cite{hu2018scai}, \cite{rattanyu2010emotion}\\
\hline
\textbf{Electrodermal} \textbf{Activity} \textbf{(EDA)} & \textbf{Wavelet:} max, mean, std, median, above zero (1 second and half second wavelet)
\textbf{Raw:} amplitude, max, min, mean
\textbf{Filtered:} amplitude, max, min, average  & A 1-second and a half-second window were used for wavelet features. Features were calculated for both the first and second derivatives of each window size. & \cite{huynh2021stressnas}, \cite{nalepa2019analysis}, \cite{ragot2017emotion}, \cite{sano2013stress}, \cite{zhao2018emotionsense}
\\
\hline
\end{tabular}}
\caption{Features from our 3 raw sources as well as definitions for those features that are less commonly used. References to prior works that used the same signal for affect inference.}
\label{tab:features}
\end{table}

\subsubsubsection{Segmentation}

We segmented the data into 1-minute windows with a 30-second overlap. We then calculated statistical features for each of the sensors, excluding steps. For each of these windows, we calculated historic features. To do so, we elongated each of the windows by 5, 30, and 60 minutes, then extracted the features with the extended window size (i.e., for each 1-minute window, we have not only the features from the 1 minute but also the features going back to these 4-time frames).

\subsubsubsection{EMA Receptivity Labels}

\begin{figure}
 \centering
\includegraphics[width=0.80\textwidth]{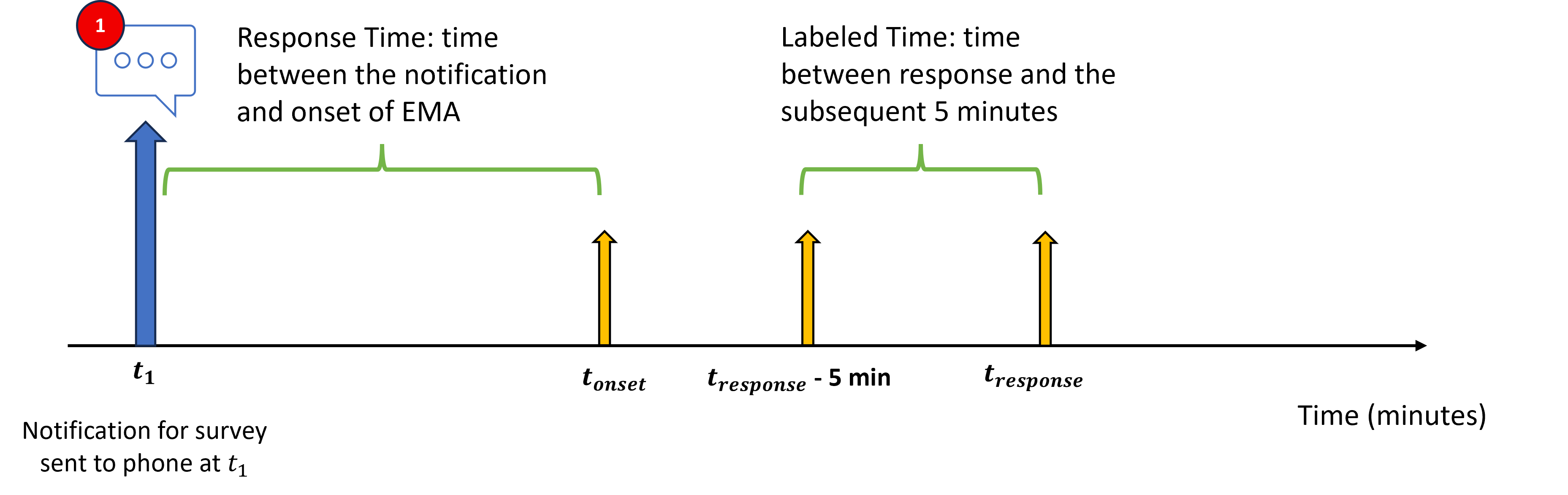}
\caption{Representation of our labeled segments for the 5-minute window. It also demonstrates how we calculate Response Time (time between notification $(t_1)$ and the start of the EMA $(t_{onset}).$ .}
\label{fig:5}
\end{figure}

Labels for receptivity were based on whether or not the user responded to the EMA and were assigned to segments based on whether it was within a specified time of the scheduled notification. By expanding the window of labeled data, we can increase the size of our labeled dataset (pseudo labeling). However, as this window increases, so does the distance between some of our time points and the corresponding label. We tested windows that are 5, 30, 60, and 120 minutes long. For instance, for the 5-minute window, if an EMA was sent at 12:00, the segments that fall between 11:55 and 12:00 would be labeled "responded" if they did respond and "no response" if they did not. We applied the same method for the affect labels; see Figure ~\ref{fig:5}. We ultimately chose 30-minute windows due to the balance between the size of the training set and the labeled points being relatively close in terms of time to the actual response (or non-response).

\subsubsubsection{Affect Labels}

Previous research on monitoring and tracking emotional states using wearables has commonly used binary or categorical affect measures to detect emotions \cite{schmidt2019wearable}. These psychological instruments often feature well-defined categorical score representations, making it easier to distinguish between emotions. The distribution of reported composite affect scores (Figure ~\ref{fig:3}) made defining an adequate categorization of the labels difficult. Most participants reported positive emotional states, which complicated setting an appropriate cutoff. Setting the cutoff at a high value would result in an imbalanced set of labels while selecting a lower value would create a balanced dataset but lack logical consistency. For instance, choosing a cutoff of 26 to distinguish between positive and negative emotions would lead to a balanced dataset. However, the range of possible responses is between 13 and 91, so a response would be considered negative even if the participant indicated relatively positive or neutral emotions.

In response to these challenges, we used recorded composite affect values as our class labels and designed our machine-learning algorithms as a regression problem. Although this method of affect inference is less commonly found in the literature, it prevents the need for arbitrary data classification. Given that the data exhibits an inherent imbalance, with less frequent occurrences of negative emotional states, using regression may still affect our ability to predict these less common negative emotions.

\subsubsection{Analysis of Relationship between Features and Receptivity/Affect }

We examined the significance of each feature in its ability to predict affect and receptivity. To do so, we conducted a repeated measures analysis of variance (ANOVA) test to assess how well each feature is related to the response class labels. Additionally, we employed a linear mixed model (LMM) to investigate the relationship between features and affect scores. We used an LMM because we utilized the constant labels rather than converting the affect score to binary or categorical like they are for receptivity. The dependent variable (affect score) was regressed on the fixed effect variable (features) while accounting for random effects (participant ID). These tests help identify any features or signals that may have significance in determining receptivity or affect. 

\subsubsection{Receptivity and Affect Model Design and Hyperparameter Tuning}

We designed machine learning models to infer EMA receptivity and affect.
There are a wide variety of machine learning algorithms utilized in affect and receptivity prediction including Random Forest (RF) \cite{ha2021wistress, huynh2021stressnas, zhao2018emotionsense}, Support Vector Machine (SVM) \cite{he2017emotion, hu2018scai, zhao2018emotionsense}, Logistic Regression (LR), K-nearest neighbors (kNN) \cite{huynh2021stressnas}, Neural Network (LSTM, RNN, CNN, etc.) \cite{huynh2021stressnas, zhao2018emotionsense}, Naive Bayes (NB) \cite{zhao2018emotionsense}, and many more \cite{zhang2020emotion}. 
Based on our sensor data, initial tests, and drawing inspiration from previous studies, especially those by Mishra et al. \cite{mishra2017investigating, mishra2021detecting}, we have selected (a) random forest (RF) for predicting emotional state and receptivity. In addition, we also used (b) neural network (NN) and (c) a baseline model. This baseline model serves as a benchmark for evaluating whether our models outperform random chance, while the Neural Network was introduced as a possible improvement on existing model implementations. Unlike the research mentioned previously, we are using physiological data rather than contextual data. These signals are sampled at higher frequencies compared to contextual data and allow for extracting more fine-grained features, making Neural Networks more feasible.
We designed personalized models to infer EMA receptivity and affect.



In order to optimize our personalized model, we selected hyperparameters using grid search method for each participant, explicitly using the GridSearchCV method defined in scikit-learn. This method uses an exhaustive search method (i.e., testing each user-defined parameter permutation). The hyperparameters tested included the number of estimators, the maximum depth of the estimator, the minimum number of samples per leaf, the minimum number of samples for split, and the maximum number of features that can be used for the split. Using a training and validation set, we selected the parameters and then applied the optimal model to our test set. The optimal set of hyperparameters differs for each participant, although the most common optimal hyperparameters chosen include 60 estimators, max depth=3, min sample leaf=2, min sample split=2, and max features=square root of the number of features.

Our NN model was structured to use three densely connected layers using a ReLu activation function at each layer. The output dimension of each layer was 256, 128, and 64, and the output layer was a densely connected layer with two output dimensions. The reasoning for an output layer of two is to define a confidence interval for our regression model.

The baseline model was built by predicting random output based on the distribution of the class labels in the training set (i.e., if 10\% of the labels are non-responses and 90\% are responses, the model would predict non-responses 10\% of the time). We can determine expected outputs for this model; our True Positive Rate should equal Pr(response in the training set) $\times$ Pr(response in the test set). The more evenly the class labels are distributed, the worse the model performs. For affect regression models, we used a normal sampling method with the mean and standard deviation based on the training set class labels.

As there are more labeled responses compared to non-responses, we considered this imbalance in the receptivity prediction model, weighting the classes based on the distribution in our training set. All the models were built using Python package scikit learn \cite{scikit-learn} or Tensorflow \cite{abadi2016tensorflow}.

\subsubsection{Model Uncertainty}

In order to determine the relationship between affect and receptivity, we have to use predictions to infer the emotional state of our participants during non-responses. And since affect is a complex and difficult-to-predict construct, we need a method for filtering our predictions based on some level of confidence. For this reason, we introduce a method for calculating uncertainty for regression using a neural network.


Determining a value for confidence for a regression model is difficult compared to a binary or categorical model. 
We can utilize a custom loss function in our neural network to estimate epistemic and aleatoric uncertainty for our regression model, where epistemic uncertainty is based on our ability to predict our class labels with the data available (affected by lack of knowledge or data), and the aleatoric uncertainty is affected by randomness that is unknown or unmeasured in the model \cite{der2009aleatory}. 

 Rather than a single predicted output, our affect prediction model outputs are two-dimensional. The first output is the predicted affect, $\mu(x)$, and the second output is the predicted variance, $ln(\sigma(x))$  (the log allows us to take the exponent to ensure a positive value for $\sigma$). Both $\mu$ and $\sigma$ are functions of our training set $x$.

\begin{equation}\label{eg:sig}
L= \bigg( \frac{1}{n} \bigg) \sum_{i=1}^{n} \Bigg( 2 \ln{\sigma(x_i)} + \bigg( \frac{y_i - \mu(x_i)}{\sigma(x_i)} \bigg)^2 \Bigg)
\end{equation}


The loss function $L$  is shown in equation ~\ref{eg:sig} and derived from the Mean Square Error (MSE) calculation and the maximum likelihood of a normal Gaussian distribution \cite{valdenegro2022deeper}. 
 The numerator of this equation is identical to the MSE loss function, where $\mu(x)$ is the predicted output of our model. Unlike the MSE loss function, we continuously update not only our predicted output $\mu$, but also the predicted variance $\sigma$. The sigma output of our model is based on error; the sigma value increases to account for higher error and decreases to account for lower error. This $\sigma$ value can be used as an uncertainty or error metric. While it is still a predicted value, it should align with how confident the model is in the $\sigma(x)$ output. The sigma value plays a crucial role in assessing the confidence of our affect predictions, given that we use predicted affect to infer emotional states during non-responses. 
 
 Consequently, to illustrate the relationship between the predicted sigma value and model uncertainty, we performed a mixed-effect model analysis using affect scores and the predicted sigma values and tested whether greater uncertainty will occur in emotional states that are less frequently represented and when the testing error is larger. As uncertainty is a measure of the model's confidence in its predictions, we can reasonably assume that predictions associated with larger testing errors would correspond to higher levels of uncertainty.

\subsubsection{Model Evaluation}

For cross-validation, we used a personalized random train-test split cross-validation method. We randomly split the data into a training and testing set using the response label (whether they responded to the EMA or not) to stratify the split. Responses and non-responses can encompass multiple segments; by grouping them together, we avoid splitting up segments from a single response or non-response. Since our response labels are unbalanced, we want to ensure that our training, validation, and test sets have a relatively even number of responses and non-responses. For the purpose of fairness, we excluded 3 participants who had a single non-response from our receptivity results.

We first normalized the training and test set independently of one another based on the participant. In total, we had around 230 features derived from the sensor signals. We reduced our feature set using Principal Component Analysis (PCA).
Our implemented PCA was set so that the number of produced components explained 99\% of the variance (48 features). This method was used for each model, excluding the Random Forest model, where we used the original normalized data as input. 


\subsection{Analyses of the Relationship between Affect and Receptivity}

We conducted two different analyses to understand the relationship between affect and receptivity better:
\begin{enumerate}

\item In order to infer emotional state during non-responses, we clustered the physiological data and then examined the makeup of the clusters. By doing so, we can assume the emotional state of different clusters and of unlabeled data points.

\item For EMAs the participants did not respond to, we used the affect prediction model described in the previous section to infer the emotional state at the time of a non-response. With these newly predicted affect scores, we can analyze the differences in the emotional state during a response and non-response.
\end{enumerate}

\subsubsection{Cluster Evaluation}

We used the most significant features (based on correlation) when predicting receptivity for our clustering analysis. To find the optimal clustering method, we tested several clustering methods, including Hierarchical and K-means clustering, with a maximum number of iterations of 300. We then calculated the silhouette score across all clusters using receptivity as our ground truth and selected our best-performing set of hyperparameters. Based on the cluster distribution, we analyzed the difference in the perceived emotional state of the participants. We calculated the average NA, PA, and receptivity rates in the clusters for each participant, and then characterized the clusters based on receptivity rates (high receptive and low receptive clusters). Next, using repeated measure ANOVA, we demonstrated the statistical difference between affect and the clusters. Given the clusters were created from physiological data, we know that the data points within each cluster are physiologically similar, so we inferred that they would also exhibit similar psychological states. This allowed us to assign affect scores to non-responsive data points within each cluster based on the labeled data points within that cluster. Unlike affect prediction, we utilized the raw NA and PA values in our evaluation as the clustering was done independently of affect scores; therefore, the lack of variance in responses did not affect the output of the clustering. These results gave us a sense of participants' perceived emotional state during non-responses. We also investigated differences in receptivity in two clusters using the chi-square test.

\subsubsection{Analysis of Receptivity and Affect Relationship}

Ideally, we would show the interaction between affect and receptivity with the data collected. However, since non-responses do not have a corresponding affect score, we designed and implemented our models for receptivity and emotional state.

After generating predictions for our test dataset, we assessed the agreement (utilizing Cohen's Kappa) and correlation (using the point biserial method) between receptivity and predicted affect, leveraging true labels at time points when affect measures were reported. A high level of agreement or correlation would suggest a strong relationship between these two constructs, thereby highlighting the potential influence each construct would have on a machine learning algorithm to predict the other construct.
%
%
We then examined the disparities between predicted affect during non-responses and reported affect during responses.  By doing so, we can establish the extent to which emotional state influences receptivity. Substantial disparities in affect between responses and non-responses suggest that participants' emotional states impact their receptivity. Consequently, a model designed to predict receptivity would indirectly include emotional state as a determinant of a participant's receptiveness. However, it is essential to acknowledge that some of these variations could be attributed to model error. As a result, we also compared predicted and reported affect during responses to investigate the significance of model error.  We then calculated and visualized the cumulative distribution of these three sets of values to illustrate the influence of affect on receptivity and the associated model error.

Finally, we explored the potential effects on reported affect that an ML-based receptivity algorithm would have on the study's outcome. Based on our receptivity model, we can estimate the difference in the reported perceived emotional state between our true findings and predicted affect during time points that would initiate an EMA.


\section{Results}

In the following section, we will discuss the results of our study, particularly the methods of evaluation that were discussed in the previous section.

\subsection{EMA Analysis - Affect and Receptivity}

\begin{figure}
 \centering
\includegraphics[width=0.65\textwidth]{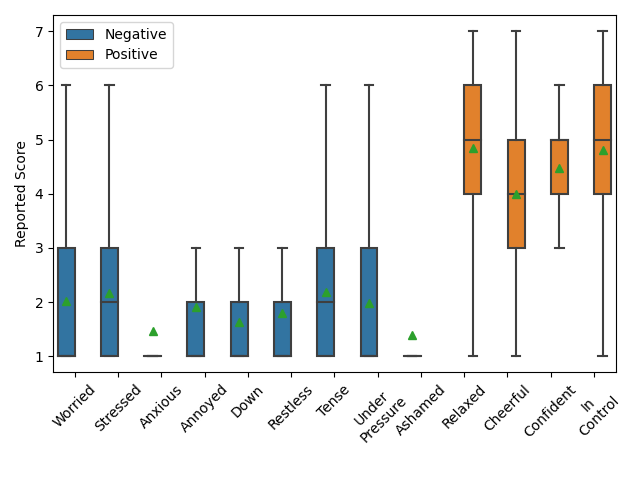}
\caption{Question Set: Includes the 13 questions we use to measure affect with their mean, standard deviation, and correlation to the final affect score. For each question, participants were asked to rate the degree they were experiencing each emotion. These 13 questions can be split into Positive Affect (PA; orange) and Negative Affect (NA; blue)}
\label{fig:EM_State}
\end{figure}

\begin{figure}
 \centering
\includegraphics[width=0.95\textwidth,trim={0cm 0 0cm 0}]{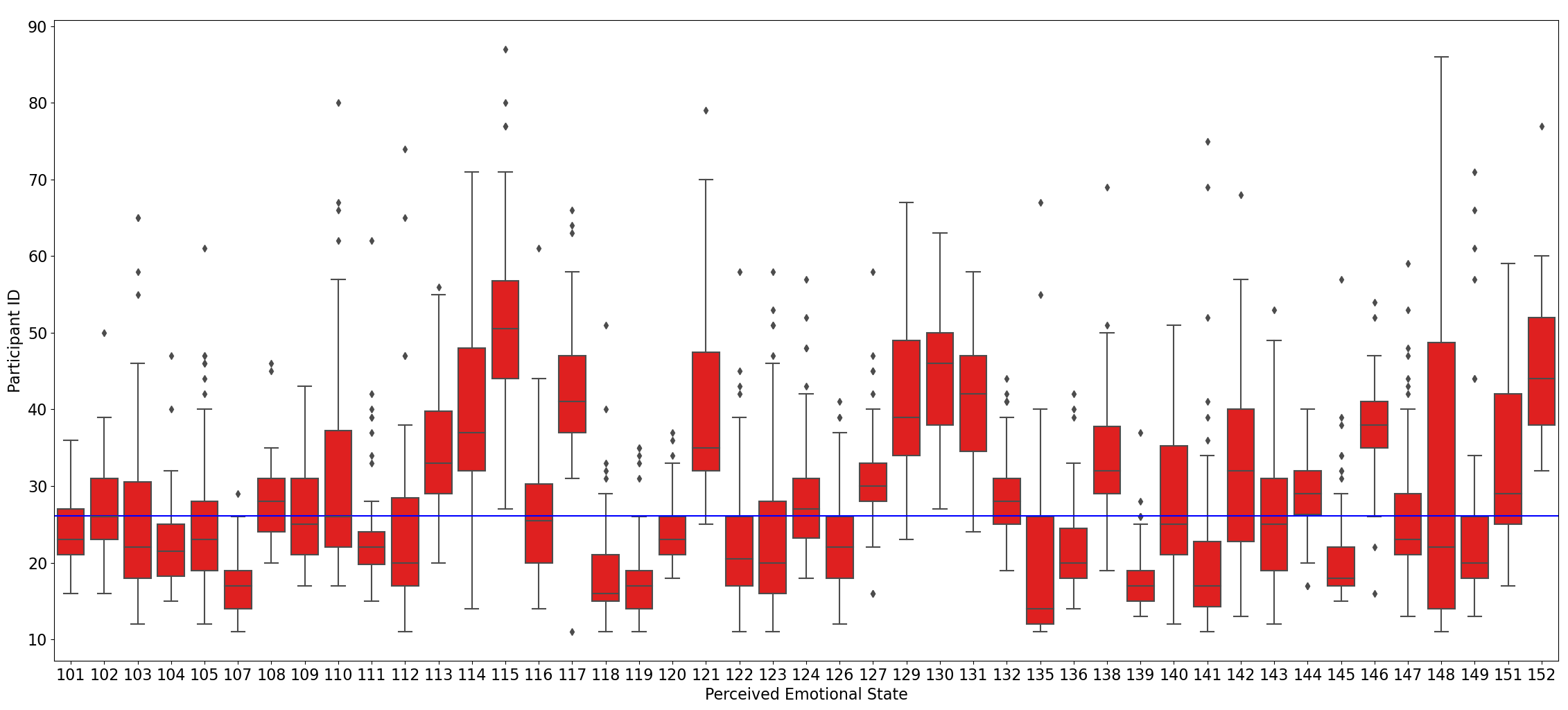}
\caption{Box plot of perceived emotional state, min is 13 (Negative) max is 91 (Positive). The average perceived Emotional State is 26.42, denoted by the blue horizontal line.}
\label{fig:3}
\end{figure}

The distribution of EMA responses can be seen in Figure ~\ref{fig:EM_State}. 
Although participants rarely indicated high negative emotion, this can be seen in Figure ~\ref{fig:3}, where we show a boxplot of each participant's composite scores. Participants' average and median reported affect were under 26, meaning that on average, the participant responded to each question with a relatively low score of 2 (on a scale between 1 and 7 where 1 indicates high positive emotion and 7 shows high negative emotion). We also investigated participants' emotional states as the study advanced and observed minimal to no variations based on their duration of enrollment or the time of day.

On average, participants responded with a 4.5 for positive affect (PA) questions and a 1.8 for negative affect (NA) questions.  This disparity in affect intensity was consistent with past research \cite{myin2001emotional}. There was a slight difference in reported affect between genders. On average, females responded with a 1.9 (stdev. = 1.08) for negative affect questions and 4.5 (stdev. = 1.3) for positive affect questions. While males responded with a 1.8 (stdev. = 0.9) for negative affect questions and 4.7 (stdev. = 1.0) for positive affect questions.

There were 3885 notifications sent to the 45 participants and 3066 responses for a receptivity rate of just under 79\%. As the study persisted, there was little to no drop-off in receptivity rates over time. This finding helped confirm that loss of engagement was not a contributing factor to receptivity. 
Most studies say that the quality receptivity rate is at 80\%. The range of response time (time between notification and initiation of the EMA Figure ~\ref{fig:5}) was between 0.5 seconds and 306 seconds. Participants responded to the notification on average in 20.9 seconds and a median response time of 8.7 seconds.  There were no responses after 306 seconds of a notification.
The reason for this fast response time is that participants were allowed 90 seconds to begin the survey, after which the survey would no longer be accessible (we had a few responses after the 90-second restriction due to software or design issues). This restriction makes it challenging to relate response times to participant affect, as other researchers have done.

We found that none of the mood responses were strongly correlated with time to respond. Across each question, we did not obtain a correlation coefficient greater than 0.03 (all correlations indicated significant confidence p < 0.05). This low correlation coefficient would indicate that the participant’s mood had little to do with how long it took the subject to initiate the EMA. Although considering the limit we put on response time, this relationship might be difficult to assume.

\subsection{Analysis of Features}

The features that we found to be the most significantly related to receptivity were ECG low frequency (1 min, momentary F(2,54) = 6.7, p<.001) and very low-frequency features (1 min, momentary F(2,54) = 4.7 p=.02 and 60 minutes, F(2,54) = 4.1 p<.001), EDA mean F(2,54) = 10.2 (p<.001) and median F(2,54) = 15.4 (p<.001), NNI (5 and 60 mins), pNNI50 (30 mins, F(2,54) = 11.2 p<.001), and max (F(2,54) = 6.3 p<.001), min F(2,54) = 3.6 (p=.009), and absolute max (F(2,54) = 6.6 p=.002) of the first and second derivatives for EDA. For the most part. ECG and EDA-related features were best at differentiating between responses and non-responses. 

When running the LMMs to determine the relationship between features and emotional state, we found a non-significant relationship between affect scores and steps or sleep features. However, heart rate was significant when predicting emotional state, particularly negative emotion. This LMM showed a significant positive relationship between heart rate and affect ($\beta$ = 0.007, p < 0.05). This underscores the significance of heart rate as a predictor of emotional state, although it does not necessarily imply that steps and sleep features lack importance in this context.

\subsection{Receptivity and Affect Models}

After processing, cleaning, and filtering out segments with confounding values, we had 1368 responses with usable physiological data. As our class labels were expanded to include segments 30 minutes before the point of response (pseudo labeling), we ended up with 13477 data points for determining affect and 17254 data points for predicting response.

\subsubsection{Model Performance}

Table ~\ref{tab:2} shows the results of our receptivity binary and affect regression models. Based on these results, there was little difference between the Random Forest and Neural Network models, although we used the Neural Network models to demonstrate the relationship between affect and receptivity in the following section. 

\begin{table}

\centering  
\resizebox{0.85\linewidth}{!}{\begin{tabular}{|c||c|c|c|c||c|}
    \hline
    
    & \multicolumn{4}{c||}{Receptivity} &  \multicolumn{1}{c|}{Affect}\\
\hline 
Model & ACC & Precision & Recall & F1 & RMSE  \\
    \hline\hline

Baseline  &   0.73 (0.001) & 0.83 (0.002) & 0.84 (0.002) & 0.83 (0.002) &   11.1 (4.3)      \\
    \hline
NN   &       0.84 (0.19) & 0.82 (0.006) & 0.85 (0.1) & 0.86 (0.2) &  7.3 (2.7)               \\

\hline
RF   &  0.83 (0.11) & 0.82 (0.15) & 0.94 (0.1) & 0.87 (0.12) &  7.5 (3.1)       \\
\hline
\end{tabular}}
\caption{Model results for predicting Receptivity (Binary) and Affect (Regression). Average evaluation metric across participants followed by the standard deviation in parenthesis.}
 \label{tab:2}
\end{table}

\subsubsection{Analyzing Uncertainty in Affect Model}

\begin{figure}
\centering
\begin{subfigure}{.5\textwidth}
  \centering
  \includegraphics[width=\textwidth]{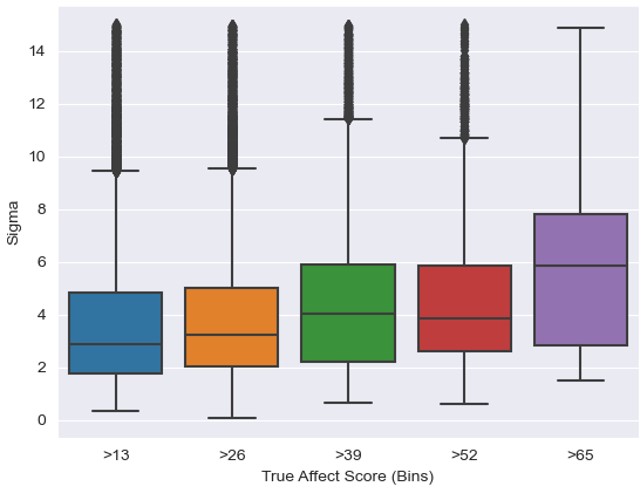}
  \caption{}
  \label{fig:6a}
\end{subfigure}%
\begin{subfigure}{.5\textwidth}
  \centering
\includegraphics[width=\textwidth]{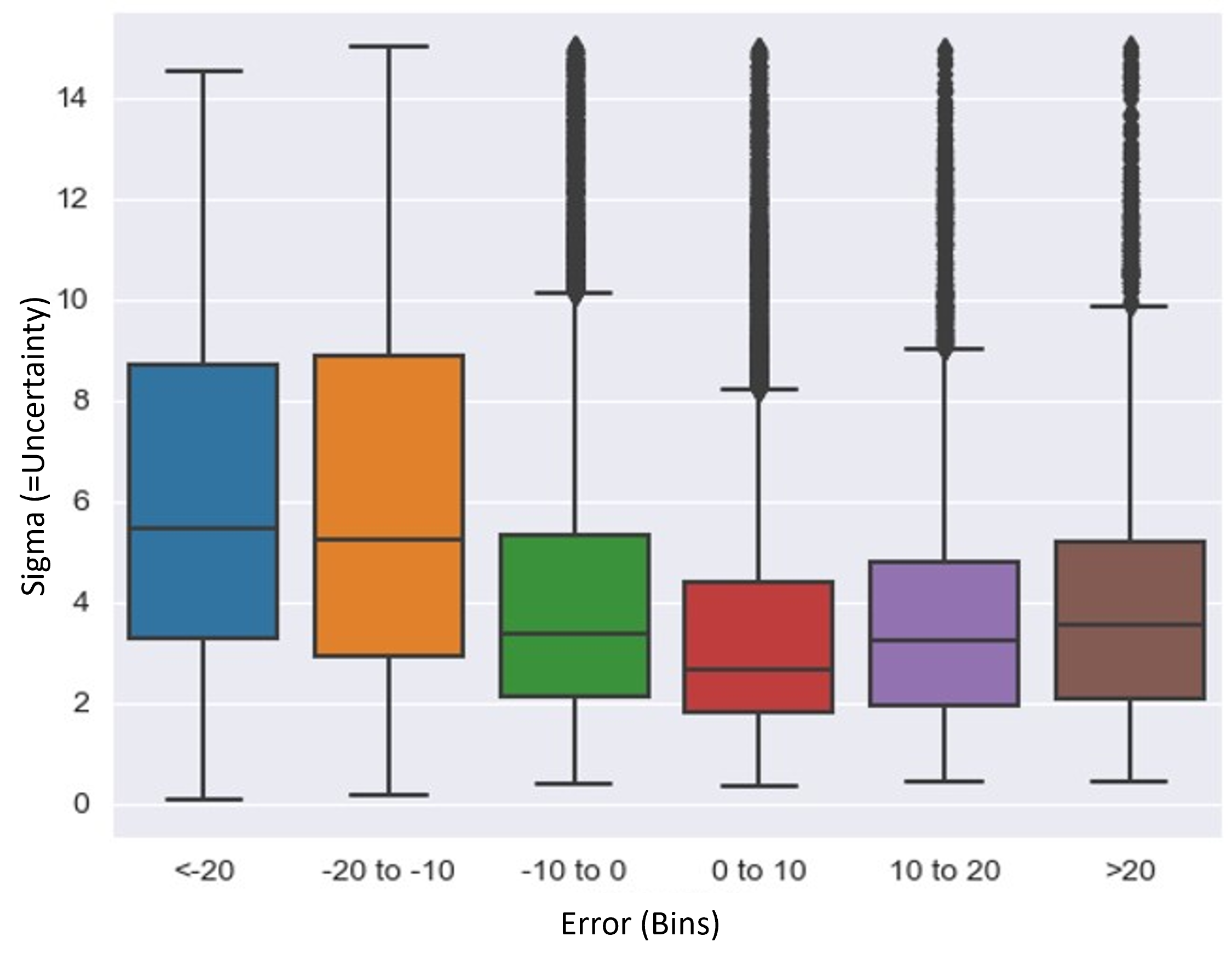}
\caption{}
\label{fig:6b}
\end{subfigure}
\caption{Box plots showing Sigma value for (a) True Labels and (b) Predicted Error}
\end{figure}

Figure ~\ref{fig:6a} helps visualize the relationship between the calculated sigma value (uncertainty) and the reported affect scores. Uncertainty should follow a pattern where class labels that are more represented in the training set should have lower uncertainty. Conversely, values less represented in the dataset should have larger uncertainty. As you can see, Figure ~\ref{fig:6a} sigma values are smaller when the reported emotional state is more positive. If you recall Figure ~\ref{fig:3}, most respondents indicated relatively low composite scores, with few participants reporting an affect score greater than 40. We also observed a statistically significant relationship between sigma and the affect scores, as defined in Figure ~\ref{fig:6a}, using a mixed effect model. In this model, we accounted for the random effect associated with participants, as indicated by the mixed linear model results (Intercept: 7.090, p < 0.001; affect score: 0.002, p = 0.046).

Figure ~\ref{fig:6b} shows the relationship between sigma and testing error, particularly that sigma values were larger when the model was further from the ground truth. This relationship is promising that our sigma value is an accurate representation of model uncertainty. Based on these two figures, we can say that the sigma value we calculated is related in some way to uncertainty. Figure ~\ref{fig:6b} shows that the majority of responses indicating an affect score of less than 39 have a sigma of less than 6. Therefore, we have chosen 6 as our cutoff for uncertainty. This cutoff filters out many of the predictions that are more likely to have higher errors since we cannot look at errors during non-responses as we have no affect label.


\subsection{Receptivity and Affect Analyses}

\subsubsection{Cluster Analysis}
\begin{figure}
 \centering
\includegraphics[width=0.5\textwidth]{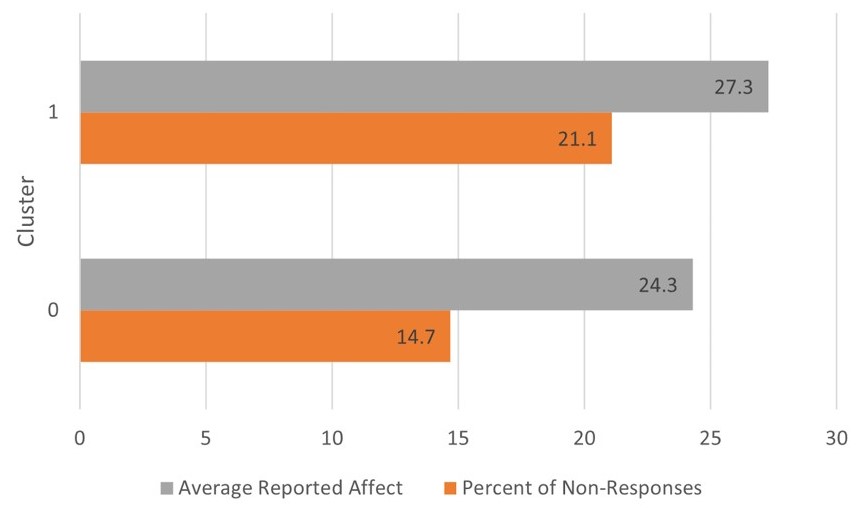}
\caption{Receptivity rates and average reported affect scores in each cluster.}
\label{fig:8}
\end{figure}

Based on the "elbow rule" of silhouette scores, we chose k-means as our clustering method with 2 clusters. We found that the distribution of receptivity was somewhat different between clusters. Cluster 0 contained a higher density of responses, with just under 15\% non-responses, while cluster 1 had a higher density of non-responses of just over 21\%. We first analyzed the overall affect scores in the two clusters, where we found the average reported affect score in cluster 1 to be more than 3 points higher than the average reported affect in cluster 0 (repeated measure ANOVA, F value 23.16, p<.001). The receptivity rates and average reported affect scores in the two clusters can be found in Figure ~\ref{fig:8}. Then we also found the distribution of receptivity was different between the two clusters using the Chi-square test of independence (test-statistic = 898.8, p<.001). These results indicate distinctions between response and affect across the cluster labels. Considering that the cluster with a higher density of non-responses (cluster 1) also had a higher average affect score (higher scores indicate more intense negative emotions or lower positive emotions), we can assume that there was a relationship between EMA receptivity and reported affect. 

\begin{figure}
 \centering
\includegraphics[width=0.5\textwidth]{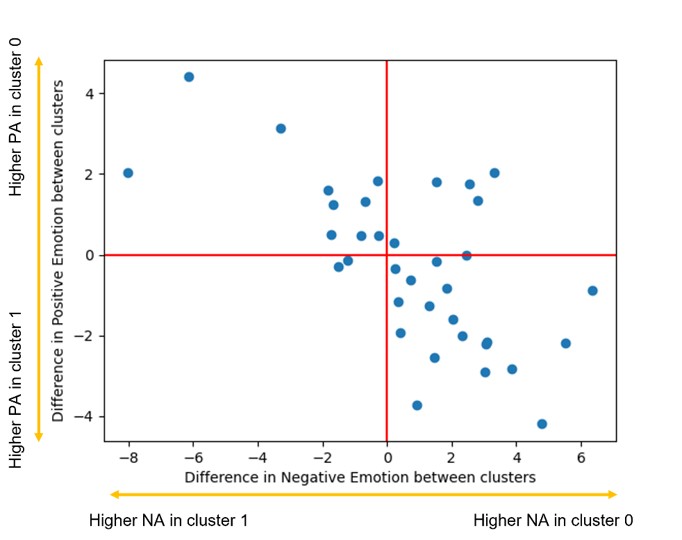}
\caption{ Each point represents a participant where the x axis denotes the difference between average NA of cluster 0 and 1, while the y axis represents the difference between average PA of clusters 0 and 1.}
\label{fig:7}
\end{figure}

Figure ~\ref{fig:7} is a scatter plot of the difference in perceived positive affect (PA) between the two clusters and the difference between negative affect (NA) between the two cluster for each participant. The results show that participants' perceived emotion was more negative regarding lower PA and higher NA in cluster 1 compared to their perceived emotional state in cluster 0. As we stated earlier, cluster 1 contains a higher percentage of non-responses compared to cluster 0, indicating cluster 1 is a better representation of a non-response. Therefore, it would appear that there is a relationship between negative perceived emotional state and receptivity. Using the cluster labels as groups, we calculated the f-score using an ANOVA test of each feature. The features that separated the two clusters were mostly calculated from the ECG signal, including minimum heart rate, low/very low-frequency, mean heart rate, CVNNI, CVSD, high frequency, and maximum heart rate (in order of f-score). Features obtained from EDA, ACC, and body temperature did not return significant p-values when calculating the f score.

\subsubsection{Relationship and Analysis between Receptivity and Affect}

\begin{figure}
 \centering
\includegraphics[width=0.85\textwidth,trim={3cm 0 3cm 0}]{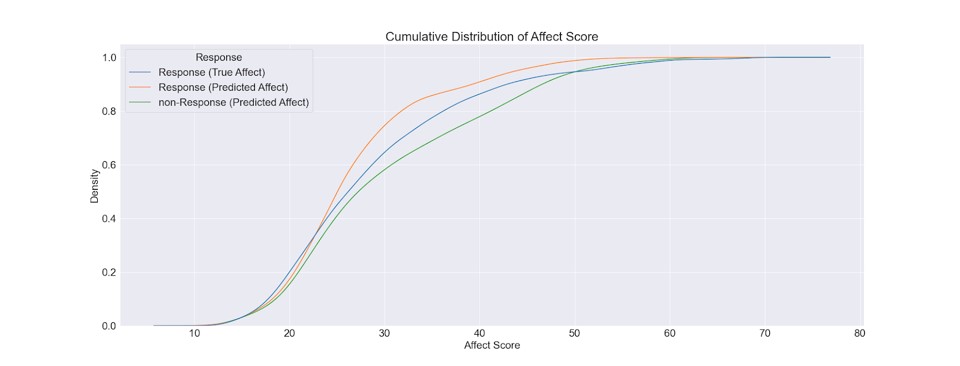}
\caption{Cumulative distribution of predicted and actual affect scores for responses and non-responses.}
\label{fig:9}
\end{figure}

Figure ~\ref{fig:9} shows the cumulative distribution of reported affect scores for responses and predicted affect scores for responses and non-responses. Based on this figure, there is a clear difference between predicted affect during non-responses and our true affect scores. While this could be a model error, we also predicted affect scores during these responses and found that our model consistently predicts lower affect values (higher positive affect). 

There was a fair amount of agreement between our affect and our binary response model with a Cohen's Kappa score of 0.33 and a correlation of 0.44. When our model predicted a response, 77\% of those responses were during times when the affect model predicted positive affect. While only 69\% of the predicted non-responses reported positive affect. This indicates that the predicted response is negatively related to affect (i.e., responses are associated with positive affect while non-responses are associated with negative affect). The reason determining the relationship between our constructs is important is because this bias can and as we show, affect the overall outcome of a study. For instance, the average predicted affect score for times that we predicted as low likelihood for a response was a full 1.5 standard deviations (1.35) or 2.01 score more than the average predicted affect for points predicted to be of high likelihood for a response. When observing just the segments where we misclassified a response (i.e., we have a true affect, but the response was misclassified as a non-response), we found that the average affect score dipped slightly from 26.1 (predicted non-response) to 25 (predicted response). This difference in affect between responses and non-responses is evidence that our receptivity model is indirectly based on affect. The standard deviation of the affect score also decreased from 11.1 (true labels) to 9.8 (true affect and predicted response) during responses. 

\begin{figure}
 \centering
\includegraphics[width=0.7\textwidth]{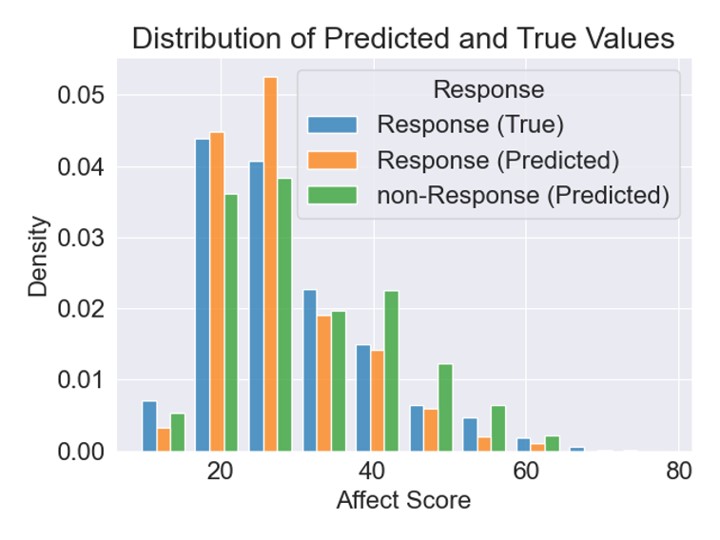}
\caption{Distribution of predicted and True Affect scores for responses and non-responses. Density is specific to the Response and non-Response.}
\label{fig:10}
\end{figure}

The average predicted affect score for a non-response was 30.9 (11.2), and the average affect score for a response was 29.3 (10.7) (True) and 27.7 (8.9) (Predicted). Predicted affect scores during non-responses were higher than reported and predicted affect scores during responses. Given that our average testing error was -1.6, we could also assume that predicted affect during these non-responses could be more negative than true predictions. The distribution of these scores can be seen in Figure ~\ref{fig:10}. In Figure ~\ref{fig:10}, all three groups' affect scores peaked at around 20-25; this is probably due to a large number of reported affect scores in this range. However, non-responses had a second peak at an affect score of 40. This bimodal distribution could indicate that our affect distribution during non-response was affected by two or more factors. Some non-responses may not be affected by their affect, but perhaps by their daily life activities (seeing a movie, spending time with family, showering, etc.). In contrast, the second peak would indicate that negative affect is related to non-responses.

\section{Discussion}

In this section, we will discuss the outcome of our study, particularly the relationship between emotional state and receptivity, what that means, how it affects our results, and how we might implement a receptivity model that removes this bias. We will also mention the limitations of this work and study.

\subsection{Principal Results}

This work aimed to understand how machine learning models, used to improve subject receptivity, can affect the outcome of a study. While we focus on the emotional state in this work, we feel as if there are many health constructs and outcomes that can be affected by these receptivity models. Improving receptivity is not a new concept, but in the realm of mHealth, it is an emerging problem. The factors influencing study adherence have been analyzed and discussed in depth in past research. One such scope is in medication adherence. Researchers have found many factors that affect medication adherence, from social, therapy, patient, or disease-related \cite{gast2019medication}. However, few have looked at the momentary factors that affect adherence to medication or a health construct, and few have had the ability to without wearable sensors and momentary assessments.

Our findings using supervised learning and clustering indicate a clear relationship between emotional state and user receptivity. The clustering method demonstrated clear differences in affect between a highly receptive cluster and a less receptive cluster. The results of supervised learning demonstrate that users are experiencing more negative emotions during non-receptive time points.  While our results showed promise for a model dedicated to predicting response, we also showed the biases in a model like this. Ideally, we would want a receptivity model completely independent of emotion. Otherwise, we are influencing the subject's responses.

Our results demonstrate that a mHealth study implementing a receptivity trigger based purely on the likelihood to respond (a model that triggers EMAs and JITIs using predicted receptivity) will bias the subject's response. In this case, the model would initiate an EMA or JITI during times of more positive emotions, therefore decreasing the overall affect score for the EMA and possibly sending the JITI during times when the intended construct is not being met. Since our ability to predict binary affect is limited with this dataset, we believe that using the affect regression and ground truth labels for responses will return the most realistic representation of affect during non-responses.


\subsection{Comparison with Prior Work}

Our findings are consistent with prior research. Much of the prompt-level studies \cite{murray2023prompt, ponnada2022contextual, rintala2020momentary} found a relationship between non-responses and higher levels of negative affect in previous prompts. While these results can give insight into what makes a participant less compliant with EMAs, they do not offer a reasonable method for using this information in real-time decision-making. Using machine learning, wearable sensors, and contextual cues allows researchers to predict non-compliance components and distribute EMAs accordingly.

Our models consistently outperformed or matched the performance of previous researchers. We achieved F1 scores ranging from 0.83 to 0.87 when predicting receptivity. In contrast, Kunzler et al. reported F1 scores of approximately 0.4 while relying solely on contextual features \cite{kunzler2019exploring}. It is important to note that these results are not directly comparable, as contextual data lacks the granularity of the data collected in our study.

Regarding affect prediction, our results present a unique challenge for comparison because we employed regression in our predictions, unlike most researchers who typically use binary or categorical labels for emotion recognition. We chose not to convert our ground truth data into binary or categorical labels due to the inherent ambiguity in setting the thresholds and the limited variance in user responses. The effectiveness of affect prediction can vary significantly depending on the specific construct of interest and the sensors and signals available. Schmidt et al. conducted a review and reported emotion recognition accuracy ranging from 40\% to 95\% using wearable sensors and signals \cite{schmidt2019wearable}. In terms of regression analysis, Tuarob et al. achieved nearly identical Root Mean Square Error (RMSE) scores (PA = 7.37, NA = 7.40) when forecasting positive and negative affect scores from PANAS using Random Forest Regression and previously collected questionnaire data \cite{tuarob2017you}.

\subsection{Limitations}

In this section, we address the limitations of our work, which can be categorized as limitations in our population, study design, data collected, and affect prediction models. 

The major concern of our population is that our results may be specific to this cohort. The population in this study was very receptive, even with 10 EMAs sent daily. This could be difficult for other researchers to implement as the frequency and complexity of the EMA are fairly burdensome. While we believe the relationship between affect and receptivity will transfer to other studies, the number of participants was relatively low (N = 45), young (age = 24.5), and had a higher distribution of females (85\%). Consequently, our results may be specific to our cohort and EMA question set, but previous studies analyzing medication adherence and prompt-level relationships between EMAs and non-responses indicate that the effect of emotional state on receptivity is common across multiple populations \cite{murray2023prompt, ponnada2022contextual, rintala2020momentary}. Further research is needed to explore the extent of this relationship between differing emotional states and receptivity across multiple populations.

One limitation caused by the study design is that we cannot examine how the loss of engagement over time affects the relationship between emotional state and receptivity. There was little to no drop-off in receptivity rates as our study progressed. This may have been due to the relatively short time frame participants were enrolled. As a result, it is difficult to explore the effect emotional state would have on EMAs in the latter part of a study when participants can be more fatigued and less engaged. In future work, we intend to study a population for an extended period of time to analyze how emotional state affects participant response rates later in the study. Ideally, this will allow us to see the rate at which responses decay, the causes, and how we might combat it. Furthermore, we believe a measure of this decay in engagement could be included in our ML-based decision-making for delivering EMAs that mitigate study fatigue, similar to how we would use model uncertainty to diversify emotional response.

Another potential study design limitation is the app and the research phone shared with participants. Carrying two phones, especially one dedicated solely to responding to EMAs, can prove burdensome for participants. Additionally, the app we designed for EMA distribution requires further usability evaluation. In future work, we aim to develop an app that can seamlessly integrate onto users' devices and assess its ease of use.

The data gathered in our study was limited to physiological features and user-defined responses. While the physiological features make up a large portion of what researchers consider important for predicting psychological constructs, the dataset lacks in sampling contextual data. Certain contextual information is imperative for recognizing emotion and improving EMA response rates that cannot be obtained using physiology, like social context. Social context can help infer the participant's emotional state and willingness to respond to an EMA or JITAI. 

Similarly, by incorporating more psychological and environmental cues (personality traits, working hours, etc.), we can better understand what to expect from our subjects regarding receptivity and affect, prior to the start of the study. Using these pre-study measurements, we can assess the type of participant enrolled. Specifically, what will be their needs regarding receiving and responding to EMAs. This will help us develop and personalize our machine learning models for affect and receptivity.

The last significant limitation of our work is the use of predicted affect labels in determining the relationship between emotional state and receptivity. We can never collect reported affect during non-responses for this or any datasets. We attempt to reduce this limitation by utilizing uncertainty to filter out less confident predictions. Nevertheless, the predicted affect is only as good as our models. The only way to curtail this limitation is by improving the affect models. While some may argue that the quality of our models need to be more robust to claim a relationship between affect and receptivity, the effects of emotional state on engagement in social and daily-life activities are well documented and consistent with our conclusion.

\section{Conclusion}

This paper presents the possibilities for bias in machine learning models to trigger surveys and interventions for participants in mHealth studies. Our results show a clear relationship between emotional state and user EMA receptivity. By designing an mHealth study using a "trigger" to improve participant response, it is imperative to consider some biases that may arise, in this case, affect. Participants were more likely to respond to an EMA during positive emotional states. If we distribute those EMAs to times when they are more likely to respond, we would further be biasing our participants' recorded emotional state. While this may not be a significant problem for less responsive populations, for the general population, this could change researchers' perception of the participant's perceived emotional state. In this paper, we did not examine other constructs that might be a factor of receptivity because affect is the focal point of this study. We are collecting both subjective and physiological data for this purpose. While this may be broad, it could be applied to any construct, particularly the intended construct of an mHealth study.

The pitfall of any mHealth study, particularly those involving psychological concepts, is the dependency on subjective user responses. The sampling rate of subjective responses will always be less than that of our physiological sensors and even some contextual cues. As our feature set becomes more and more comprehensive, our labeled data remains relatively sparse. Considering that our proposed trigger considers factors beyond receptivity, it would likely have lower receptivity rates compared to triggers solely based on receptivity. 
However, the importance of even a minimal increase in a user's adherence or engagement to a study can drastically improve researchers’ understanding of the health construct.

The models discussed in this paper have mostly proposed single-objective optimization functions that try to optimize based on whether the model thinks a user will respond to an EMA. In future work, we propose a multi-objective optimization function for triggering EMAs and JITAIs based on the likelihood to respond and an active-learning measurement of the health construct. This multi-objective function would base the timing of the EMAs on two separate objectives, include receptivity and model uncertainty. By initiating EMAs or JITAIs based on these two objectives, we can obtain an expected response that is more diverse in terms of affect. We hope that the work we have presented in this paper can be used to enhance further our communication and ability to gain knowledge from subjects.

\bibliographystyle{unsrt}  

\bibliography{references}  






\end{document}